# Methane and n-hexane ignition in a newly developed diaphragmless shock tube

Janardhanraj Subburaj[1*], Touqeer Anwar Kashif[1], Aamir Farooq[1*]

[1] *Clean Combustion Research Center, Physical Sciences and Engineering Division, King Abdullah University of Science and Technology (KAUST), Thuwal 23955, Kingdom of Saudi Arabia*

*Corresponding author email: janardhanraj.subburaj@kaust.edu.sa, aamir.farooq@kaust.edu.sa

**Abstract**
Shock tubes have been routinely used to generate reliable chemical kinetic data for gas-phase chemistry. The conventional diaphragm-rupture mode for shock tube operation presents many challenges that may ultimately affect the quality of chemical kinetics data. Numerous diaphragmless concepts have been developed to overcome the drawbacks of using diaphragms. Most of these diaphragmless designs require significant alterations in the driver section of the shock tube and, in some cases, fail to match the performance of the diaphragm-mode of operation. In the present work, an existing diaphragm-type shock tube is retrofitted with a fast-acting valve, and the performance of the diaphragmless shock tube is evaluated for investigating the ignition of methane and n-hexane. The diaphragmless shock tube reported here presents many advantages, such as eliminating the use of diaphragms, avoiding substantial manual effort during experiments, automating the shock tube facility, having good control over driver conditions, and obtaining good repeatability for reliable gas-phase chemical kinetic studies. Ignition delay time measurements have been performed in the diaphragmless shock tube for three methane mixtures and two n-hexane mixtures at $P_5$ = 10 - 20 bar and $T_5$ = 738 - 1537 K. The results obtained for fuel-rich, fuel-lean, and oxygen-rich (undiluted) mixtures show very good agreement with previously reported experimental data and literature kinetic models (AramcoMech 3.0 [1] for methane and Zhang et al. mechanism [2] for n-hexane). The study presents an easy and simple method to upgrade conventional shock tubes to a diaphragmless mode of operation and opens new possibilities for reliable chemical kinetics investigations.

**Keywords**: Methane oxidation; n-Hexane oxidation, Diaphragmless shock tube; Fast-acting valve; Ignition delay time.

# 1. Introduction

Shock tubes have proven to be ideal reactors to study a variety of gas-phase chemical reactions due to their ability to produce a wide range of temperature and pressure [3]. The conventional mode of shock tube operation uses a frangible diaphragm for shock initiation. The diaphragms used in shock tubes require strict quality control practices to ensure consistency in the material standard and appropriate manufacturing techniques that follow precise tolerances during machining. Without these practices, the reproducibility of the diaphragm rupture process cannot be ensured resulting in shot-to-shot variation in the shockwave conditions produced in the shock tube. Diaphragm rupture results in tiny broken fragments inside the shock tube and the debris affect the quality of data in future experiments. The choice of material and thickness of the diaphragm has a significant effect on the nature and time scale of rupture. Thicker diaphragms are not preferred as they resist opening at the hinge line. Also, the manual effort spent replacing the diaphragms after every experiment is substantial. These shortfalls of using diaphragms in a shock tube have prompted researchers to explore alternative methods of diaphragmless operation.

Several diaphragmless concepts have been suggested that include piston-cylinder configurations, bellow-actuated drivers and fast-acting electro-pneumatic valves [4]. Sikes et al. [5] highlighted the applications of different diaphragmless shock tubes for chemical kinetics studies. The actuating mechanisms of the diaphragmless shock tubes are generally either inside or outside of the driver section. Most of these diaphragmless shock tubes have specially designed drivers that significantly alter the flow path from the driver to the driven section. While many of the diaphragmless shock tubes operate at low pressures, high-pressure diaphragmless shock tubes are limited to miniature scales. Therefore, a diaphragmless method with minimal design modification and capable of operating at high pressures would be favorable for reliable chemical kinetic studies.

One of the earliest reports demonstrating the use of a diaphragmless shock tube for chemical reactions utilized a piston-cylinder arrangement operated by magnetic valves in a 40 mm diameter shock tube, and operated at a maximum driver pressure of 9 bar [6]. A similar piston-cylinder arrangement was adopted by Yamauchi and co-workers [7] in a 30 mm shock tube, and operated at a maximum driver pressure of 20 bar for OH and SiH emission spectra studies. This facility was coupled with Atomic Resonance Absorption Spectrometry (ARAS) to detect N atoms [8] as well as to determine reaction rate coefficients at high temperatures up to 3197 K [9]. Hurst and Bauer [10] adopted Yamauchi et al.'s design for a shock tube cross-section of 62 mm by 44 mm, albeit at low driver pressures of about 21 bar, to study the dissociation of ethane and the condensation of iron atoms. Matsui et al. [11] studied several chemical reactions in a diaphragmless piston-actuated shock tube with excimer laser photolysis and ARAS.

Bellow-actuated diaphragmless shock tubes were initially used to improve chemical kinetic data in the low-pressure regime in a 71 mm shock tube, as demonstrated by Tranter and co-workers [12]. This diaphragmless shock tube was also coupled with a TOF-MS (Time-of-flight Mass Spectrometer) to perform signal averaging over multiple experiments to study thermal dissociation of fluoro-ethane [13]. The Brown Shock Tube, based on bellow actuation, had a driven section diameter of 102 mm and was demonstrated for pyrolysis studies of isobutyl nitrite and cyclohexene [14]. A bellow-actuated diaphragmless shock tube designed by McGivern et al. [15] had a vertical driver section attached to a 31.8 mm horizontal driven section. The HRRST (high repetition rate shock tube) was designed to generate reaction conditions of T > 600 K and P < 100 bar at a cycle rate of up to 4 Hz in a 6.35 mm shock tube [16]. PI-TOF-MS (Photo Ionization Time-of-flight Mass Spectrometer) was used in this facility to study the pyrolysis of dimethyl ether [17]. A modified solenoid driver was built for a high-repetition miniature shock tube with an internal diameter of 12.5 mm [18]. This miniature diaphragmless shock tube has been used for studying 1,1,1-trifluoroethane dissociation [19], and ignition delay of fuels like toluene, n-heptane, and iso-octane fuel blends [20]. The ignition delay time data has also been correlated with those obtained using the regular shock tube data [21]. A similar shock tube with a diameter of 8 mm was utilized for pyrolysis studies using photoelectron/photoion coincidence spectroscopy [22]. A new solenoid actuated driver valve design was proposed recently to improve the longevity of the miniature diaphragmless shock tubes [23].

Diaphragmless shock tubes that employ a piston-cylinder arrangement or bellow-based actuation have modified driver sections that generally have large cross-sectional areas compared to the driven section. The hardware responsible for actuation occupies a large volume of the driver section, thus making the entire system bulky and altering the flow path from the driver to the driven section. For shock tubes with diameters greater than 30 mm, the operating driver pressure range is restricted to less than 20 bar. Higher driver pressures are used in miniature diaphragmless shock tubes. Therefore, a fast-acting valve with a wide operating pressure range, compact design, and minimal flow obstruction is desirable.

Commercially procured fast-acting valves have been used in diaphragmless shock tubes for pressure sensor calibration [24-27]. The performance of these fast-acting valves has been reported previously for driver pressure up to ~ 15 bar in terms of shock formation, repeatability, valve efficiency, and ease of operation [27]. These valves have not been used for chemical kinetic studies, to the best of our knowledge, and testing their performance in chemically reacting environments will open up new research directions.

The main aim of the present study is to evaluate the performance of a newly developed diaphragmless shock tube for gas-phase chemistry. An existing diaphragm-type shock tube was retrofitted with a commercial fast-acting valve with no major alterations to the driver and driven sections. The range and repeatability of shock conditions were analyzed and compared with the diaphragm-mode of operation. Ignition studies were performed in methane and n-hexane mixtures of different compositions with high fuel concentrations and low dilution. Experimental results are compared with chemical mechanisms that predict the ignition behavior of methane and n-hexane mixtures.

## 2. Experimental methodology

### 2.1. Experimental setup

The present work was carried out in the high-pressure shock tube (HPST) facility at KAUST. The shock tube has an internal diameter of 101.6 mm and a driven section length of 6.6 m. The driver section is modular and its length can be varied from 2.2 m to a maximum of 6.6 m. The diaphragm section in the shock tube supports both single- and double-diaphragm modes of operation. In the double-diaphragm technique, a small buffer volume is employed between the primary and secondary diaphragm, which is filled to a pressure about half of the driver pressure. After the driver section is filled to the required pressure, the buffer volume is evacuated to initiate the rupture of the primary and secondary diaphragms for shock wave production. Aluminum diaphragms are scored with a cross-shaped V-groove to obtain the rupture along a preferential plane.

The fast-acting valve used in the study was procured from ISTA Pneumatics Inc., Russia (Model: KB-80-100). The valve is operated electro-pneumatically with a pilot pressure of 8 bar. The inlet and exit diameter of the valve is 80 mm and it has an overall length of about 310 mm. Figure 1a shows the schematic diagram of the internal flow path and the major components of the valve. A sliding cap is the main moving element that is responsible for opening or blocking the pathway between the inlet and outlet. Initially, a small volume behind the sliding cap is filled to move it forward and block the driven section from the driver section. Once the driver and driven gases are filled in the respective sections, the actuating volume behind the sliding cap is quickly exhausted through an actuating valve. The cap retracts, opening the pathway between the driver and driven sections. The front face of the cap has a conical shape to streamline the flow of the driver gas into the driven section. There are bound to be losses due to the flow inside the valve that affect the opening time and the shock formation process. Nevertheless, the design of this valve facilitates direct mounting in place of a diaphragm station of a regular shock tube. The valve can operate up to a maximum differential pressure (difference between the pressure at inlet and outlet) of 100 bar. The opening time of the valve is about 7 milliseconds and previous studies using a similar fast-acting valve have shown satisfactory shock formation in a shock tube with a driven section length of 4 m [26]. This valve is different from the one used by Amer at al. (see Figure S1 in SM) with respect to the operational pressure range and the internal design (which can have significant impact on the shockwave production and flow features in the shock tube). Therefore, it is necessary to understand the shock formation ability of the valve used here and to characterize its performance over the desired operational range. These results have been discussed in detail in section 3.1. Two separate sections were fabricated to provide a smooth transition from 80 mm diameter of the valve to 101.6 mm diameter of the shock tube at the inlet and outlet of the fast-acting valve to avoid abrupt changes in the diameter. A schematic of the fast-acting valve mounted between the driver and driven section of the shock tube is shown in Fig. 1b. Acetone and tert-butyl-hydroperoxide were used to clean the shock tube, and high-temperature oxygen shocks were performed before reactive experiments with a particular mixture. Fuel mixtures were prepared in a 40 L magnetically-stirred mixing vessel and left for at least four hours before starting the experiments. Gases (oxygen and argon) with a purity of 99.999% were procured from Air Liquide and n-hexane was purchased from Sigma-Aldrich. The detailed procedure for sample preparation has been described earlier [28].

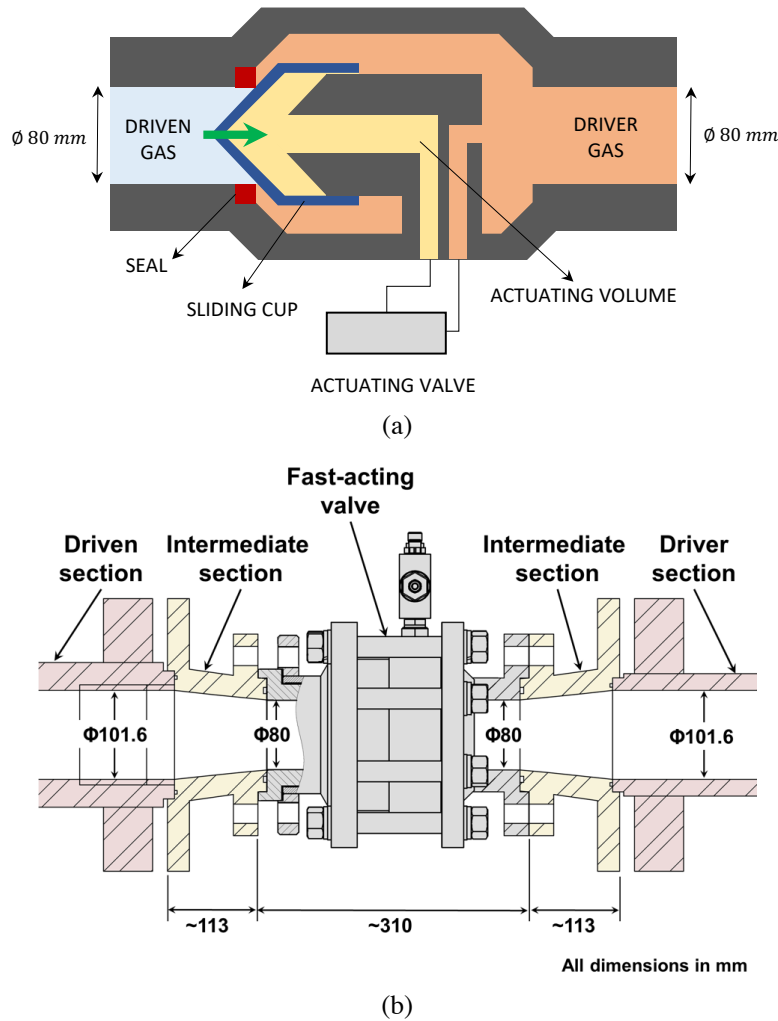

Figure 1. (a) Schematic diagram showing the internal flow path and major components of the fast-acting valve (Courtesy: US Patent US7232152B2). The green arrow indicates the retraction direction of the sliding cup upon

actuation. (b) Schematic diagram showing the fast-acting valve mounted between the driver and driven sections of the shock tube using intermediate sections that provide a transition from diameter of 101.6 mm to 80 mm.

## 2.2. Ignition delay measurements

Six piezoelectric transducers (Model: 113B26, PCB Piezotronics, USA) were flush-mounted on the sidewall of the driven section, equally spaced over 3.4 meters from the shock tube end wall. Incident shock speed was measured by the time-of-flight method, and the attenuation of incident shock was estimated for each experiment. Thermodynamic conditions behind the reflected shock were estimated from the incident shock speed, thermodynamic parameters of the shock-heated gas, and the shock jump relations using a MATLAB code [29]. The average *dP/dt* in the reflected shock region for the experiments in the diaphragmless mode was about 2.5 %/ms which is similar to the *dP/dt* observed in diphragm mode of operation in this facility. Additionally, OH* chemiluminescence was monitored at the end-wall through a port with a sapphire window using a PDA36A photo-detector. A piezoelectric transducer (Model: 603B1, Kistler, USA) was flush-mounted on the sidewall of the driven section at a distance of 1 cm from the end wall to record pressure traces for ignition delay time (IDT) measurements. Ignition delay time is defined as the time elapsed between the arrival of the reflected shock and the onset of ignition at the measurement location. Figure 2 shows typical pressure and OH* signals for IDT measurement. Due to the low vapor pressure of liquid fuels (e.g., n-hexane), there is a possibility of condensation in the post-shock region. The film thickness is on the order of microns at these timescales [30], and the mass of the fuel condensed on the walls is a small fraction of the total fuel mass used in this study (calculations shown in SM). Therefore, IDT measurements in the present operating conditions are not affected by any potential condensation of the fuel.

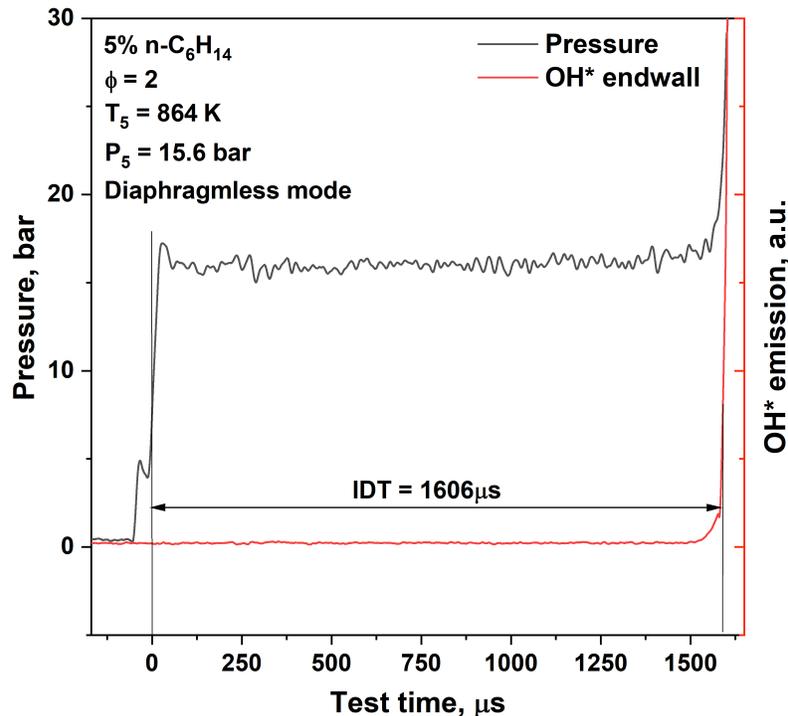

Figure 2. Representative pressure and OH* profiles for an IDT measurement in the diaphragmless mode of operation.

## 2.3. Fuel mixtures

Two different fuels have been considered as test cases to evaluate the performance of the diaphragmless shock tube. Table 1 summarizes the compositions of the mixtures used in the study and the corresponding pressure conditions for the experiments. Ignition delay characteristics and kinetic modeling of highly diluted methane fuels have been well investigated. Recently, shock tube studies were performed in oxygen-rich methane mixtures at high pressures (see Supplementary Material). Two stoichiometric methane mixtures, M1 and M3, with argon and nitrogen as diluents are studied at 10 bar and 20 bar, respectively. The effect of oxygen enrichment at 10 bar is studied using the mixture M2. Additionally, as a representative negative temperature coefficient (NTC) fuel, n-hexane mixtures have also been tested in the present work. IDT measurements in n-hexane mixtures using a diaphragm-type shock tube have been reported recently [31]. Two mixtures of n-hexane with different diluents are examined as well as a fuel-rich mixture with argon as the diluent.

Table 1: Mixture compositions and pressure conditions used in the present study.

| Fuel | Mixture | Composition | $\phi$ | $P_5$ (bar) |
|---|---|---|---|---|
| Methane | M1 | Fuel: 9.50%, $O_2$: 19.00%, Ar: 71.50% | 1 | 10 |
| Methane | M2 | Fuel: 9.50%, $O_2$: 90.50% | 0.2 | 10 |
| Methane | M3 | Fuel: 9.50%, $O_2$: 19.00%, Ar: 71.50% | 1 | 20 |
| n-Hexane | H1 | Fuel: 2.20%, $O_2$: 20.50%, $N_2$: 77.30% | 1 | 15 |
| n-Hexane | H2 | Fuel: 5.00%, $O_2$: 23.75%, Ar: 71.25% | 2 | 15 |

## 3. Results and Discussion

### 3.1. Performance of the diaphragmless shock tube

The performance of the diaphragmless shock tube was examined in non-reactive experiments before performing IDT measurements. Experiments were performed using helium at different driver gas pressures ($P_4$) and argon in the driven section ($P_1$ = 450 Torr). Although the valve can operate up to a pressure of 100 bar, the maximum pressure used in the driver section was 723 psi (~ 50 bar), limited by the operating range of the pipe fittings used for pneumatic connections. Experiments were performed for pressure ratios $P_{41}$, defined as the ratio of $P_4$ and $P_1$, in the range of 27.9 - 83.1. Experiments were also performed in the diaphragm-mode of operation at similar driver and driven conditions for comparison. Figure 3 shows a plot of the shock Mach number ($M_S$) at different pressure ratios $P_{41}$. The experimental trend obtained for diaphragmless mode of operation is similar to that obtained using diaphragms and the ideal shock tube relation (assumes an inviscid, adiabatic, and one-dimensional flow). The longer opening time of the fast-acting valve (~ 7 ms) compared to the typical rupture time scale of metal diaphragms (≤ 1 millisecond) results in lower shock Mach numbers obtained in the diaphragmless mode of operation. Therefore, a larger $P_{41}$ is required in the diaphragmless shock tube to obtain the same performance of a diaphragm-mode of operation.

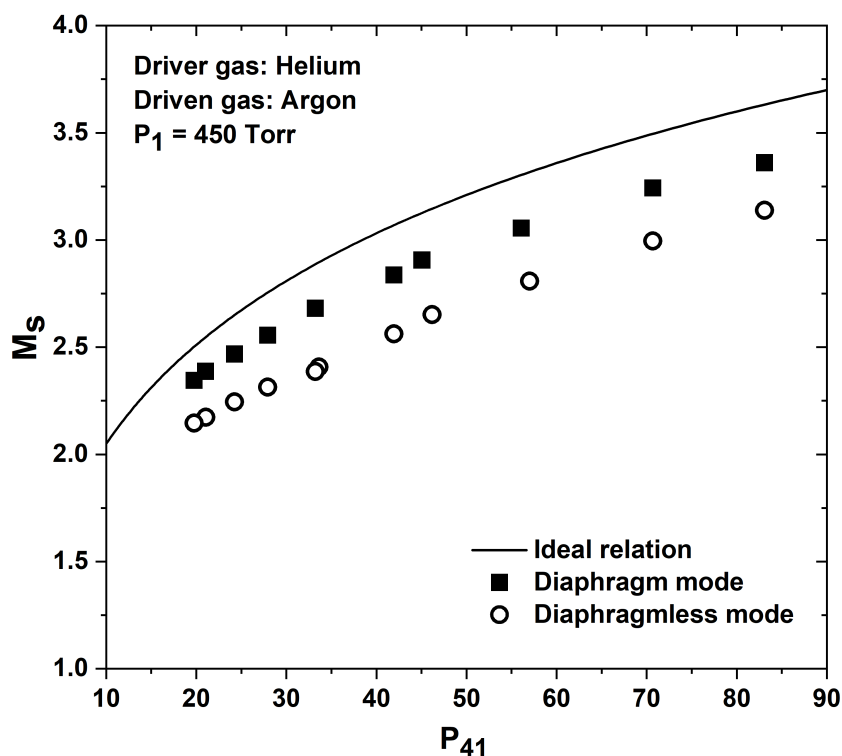

Figure 3. Comparison between the shock Mach numbers ($M_S$) obtained at different $P_{41}$ in the diaphragmless and diaphragm-mode of operations in the HPST.

The repeatability of the diaphragmless shock tube is compared with the conventional single-diaphragm and double-diaphragm mode of shock tube operation. The reproducibility of thermodynamic conditions in the reflected shock region is critical in chemical kinetics studies using shock tubes. Experiments were performed for a fixed value of $P_{41}$ and $T_5$ was calculated using the measured incident shock velocities. Initially, experiments were performed using the single-diaphragm mode of operation and the diaphragm rupture pressure ($P_4$) was fixed for the remaining experiments to characterize repeatability. Care was taken to ensure that diaphragms were manufactured from the same aluminum sheet and had identical groove depths. Figure 4 shows the values of $T_5$ obtained during the repeatability tests. Temperature values obtained in the double-diaphragm mode are less than those obtained using the single-diaphragm mode as the driver pressure expands into the buffer volume before the second diaphragm ruptures. Therefore, the rupture pressure in the case of the double-diaphragm mode is slightly less than the initial set pressure $P_4$. In the case of the diaphragmless shock tube, the driver pressure can be fixed before operating the valve and this feature is an advantage over the conventional methods of operation. $T_5$ values obtained using the diaphragmless shock tube are significantly lower than the diaphragm-modes of operation at the same $P_{41}$ as the shock Mach number is less due to the longer opening times. Standard deviations in $T_5$ values for the five different experiments show that the repeatability of the diaphragmless shock tube is similar to the diaphragm-mode of operation when utmost care is taken to replicate the manufacturing of diaphragms. Since the scope of the present work is to demonstrate the use of the diaphragmless shock tube for chemical kinetic studies, the repeatability tests were limited to five shots. According to the manufacturer, the fast-acting valves can function up to five million cycles without major issues. Nevertheless, extensive repeatability tests of post-shock conditions will be carried out in future.

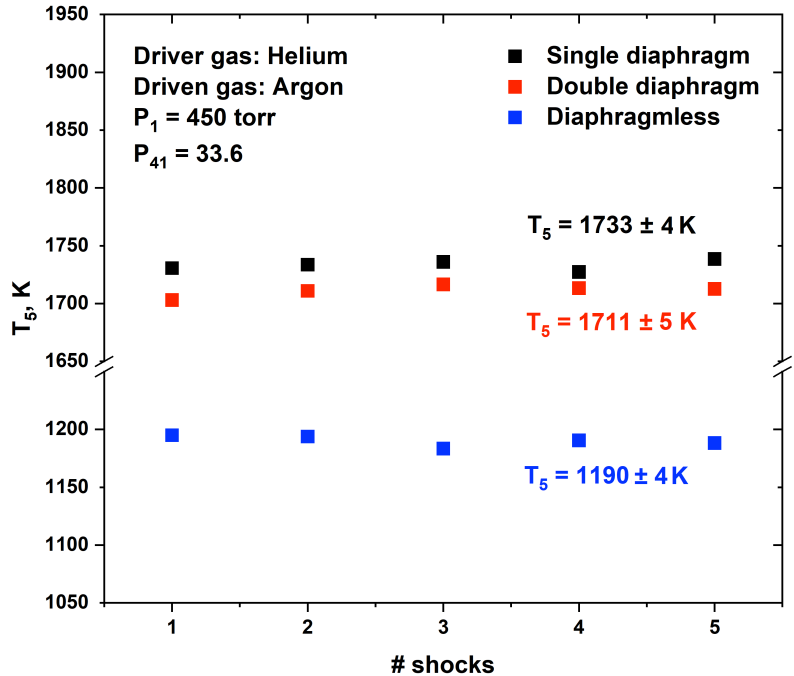

Figure 4. Reflected shock temperature $T_5$ for five experiments performed at the same $P_{41}$ using single diaphragm-mode, double diaphragm-mode and diaphragmless-mode of shock tube operation. The values are represented as $\mu \pm \sigma$, where $\mu$ is the mean and $\sigma$ is the standard deviation.

A sample shock velocity measurement, demonstrating the attenuation rates, is shown in Figure 5 for the diaphragmless mode of operation. The driven gas for this particular experiment was mixture H2 (see Table 1). In the plot, the incident shock wave travels from right to left where the distance of $x = 0$ denotes the end wall location. Incident shock velocity was extrapolated to the end wall and the obtained value is 634.9 m/s. The velocity error, in this particular case, is $\pm 0.2\%$ and the attenuation rate is about 0.41%/m. The attenuation rate is comparable to those obtained in the diaphragm mode of operation as reported in literature [32]. Pressure profiles recorded near the end wall for a reactive mixture are compared in Figure 6 for the diaphragmless and diaphragm-mode of operation at similar reflected shock conditions. Measured IDTs have a reasonably good agreement considering the IDT measurement uncertainty of $\pm 20\%$ [33]. Driver gas tailoring is employed for the experimental conditions shown in Figure 6, and the test-time, estimated using an inhouse WENO shock tube code, is about 1.8 milliseconds (shown in SM Figure S2). The gradual pressure increase prior to the main ignition is possibly due to the coupled effect of *dP/dT* and heat release due to exothermic reactions. For high fuel percentage of 9.5%, the heat release prior to ignition is expected to be quite high. Similar pressure rise prior to ignition has been observed in other pressure profiles obtained for mixture M1 (shown in SM Figure S3).

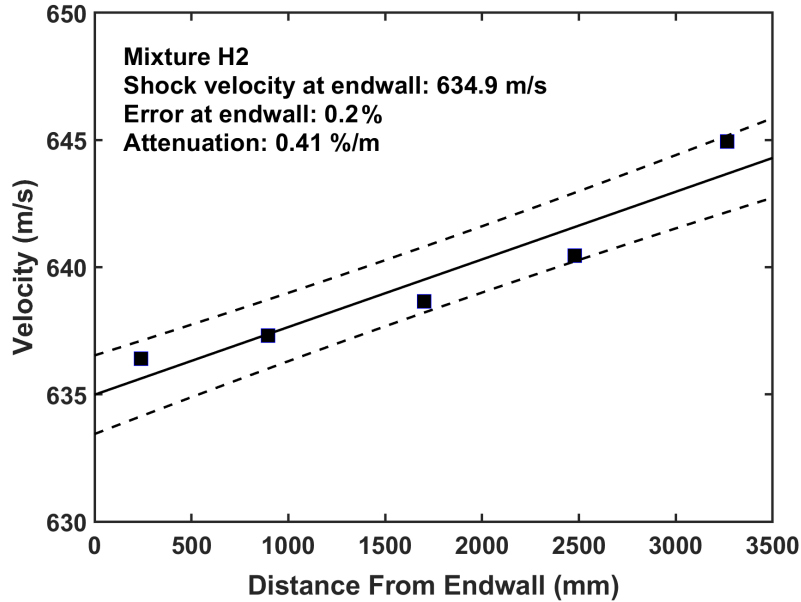

Figure 5. A sample incident shock velocity measurement for the diaphragmless mode of operation.

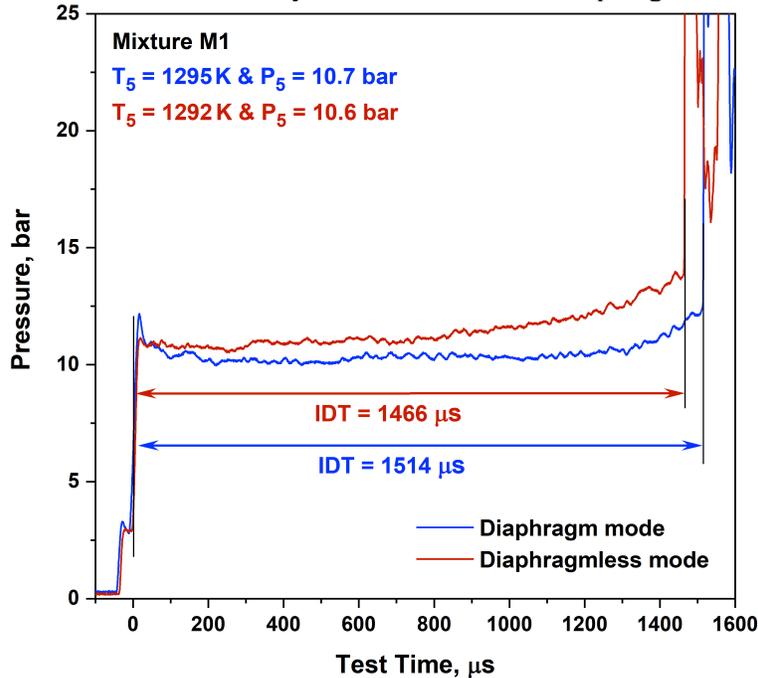

Figure 6. Comparison between the pressure signals obtained in reactive experiments for two similar reflected shock conditions in the diaphragmless and diaphragm-mode of operations.

The difference between ideal and experimentally obtained $P_5$ values is investigated for both modes of operation. Figure 7(a) shows that the difference between the ideal and experimental $P_5$ is almost constant with $T_5$ in both the diaphragm and diaphragmless modes. There is a larger drop in experimental $P_5$ (with respect to the ideal value) in the diaphragmless mode since a higher $P_{41}$ is required in the diaphragmless-mode to obtain the same $P_5$. The normalized drop in post-shock pressure is plotted against the normalized post-shock temperature for two fuel mixtures in Figure 7(b). The performance of the fast-acting valve at $P_5$ = 10 bar and 20 bar seems to be similar in terms of the drop in $P_5$. Since the focus of the present work is on IDT measurements, a more detailed analysis of the valve performance at different post-shock pressures will be presented in future.

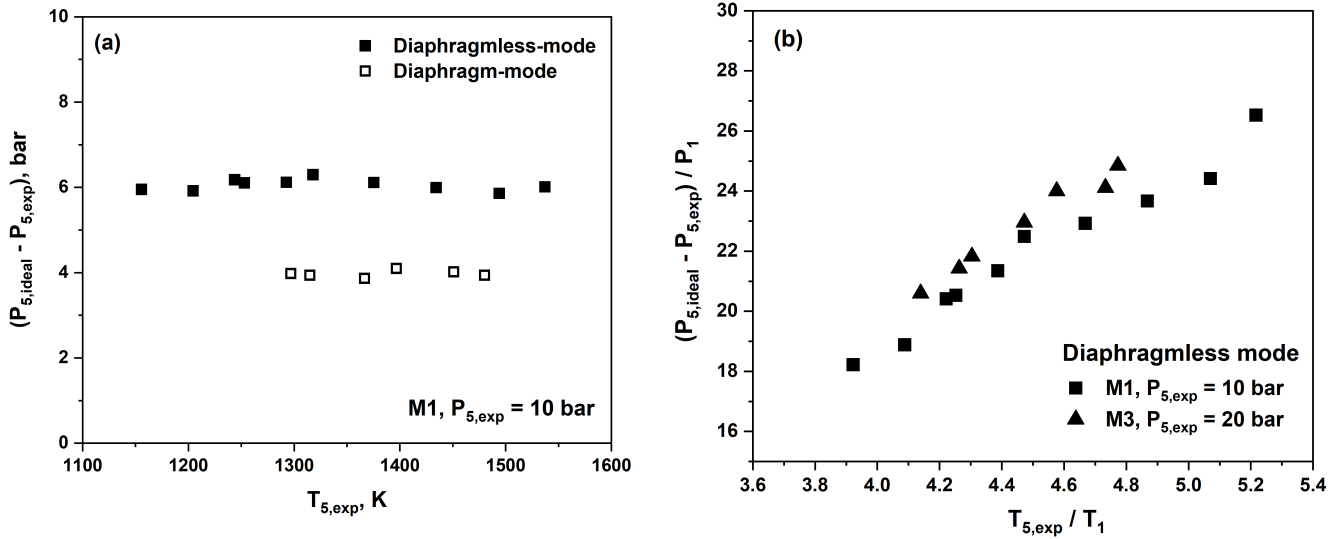

Figure 7. (a) Difference between ideal and experimentally obtained $P_5$ over a range of $T_5$ for diaphragm- and diaphragmless-mode of operation. The post-shock pressure is 10 bar and the mixture used is M1. (b) Normalized drop in $P_5$ as a function of normalized $T_5$ for two post-shock pressures (10 bar and 20 bar) in the diaphragmless mode of operation.

### 3.2. IDTs of methane mixtures

Figure 8 shows IDT measurements performed for methane mixtures in the two modes of shock tube operation. Experimental results are also compared with the predictions of AramcoMech 3.0. The simulations were performed in ANSYS Chemkin-Pro using a closed homogeneous batch reactor with a $dP/dt = 2.5\%$/ms. For the stoichiometric mixture diluted in argon (M1), there is a reasonably good agreement between the experiments in diaphragmless and diaphragm-mode of operation and they match the trend predicted by the simulations (see Figure 8a). Experimental results from Kashif et al. [34] and Burke et al. [35] have been used for comparison. In the case of oxygen-enriched mixture M2, the performance of the diaphragmless shock tube matches the diaphragm-mode of operation but the simulations predict longer IDTs at lower temperatures as compared to the experiments (see Figure 8b). Figure 8c compares IDT measurements for mixture M3. Experimental results from Kashif et al. [34], Burke et al.[35] and Heufer et al.[36] have been used for comparison. The experimental trend is similar to that obtained in the diaphragm modes of operation using the same fuel-mixture ratio with nitrogen as the diluent. The effect of diluent on the chemical behavior of methane, as predicted by AramcoMech 3.0, is well-captured by the experimental IDTs.

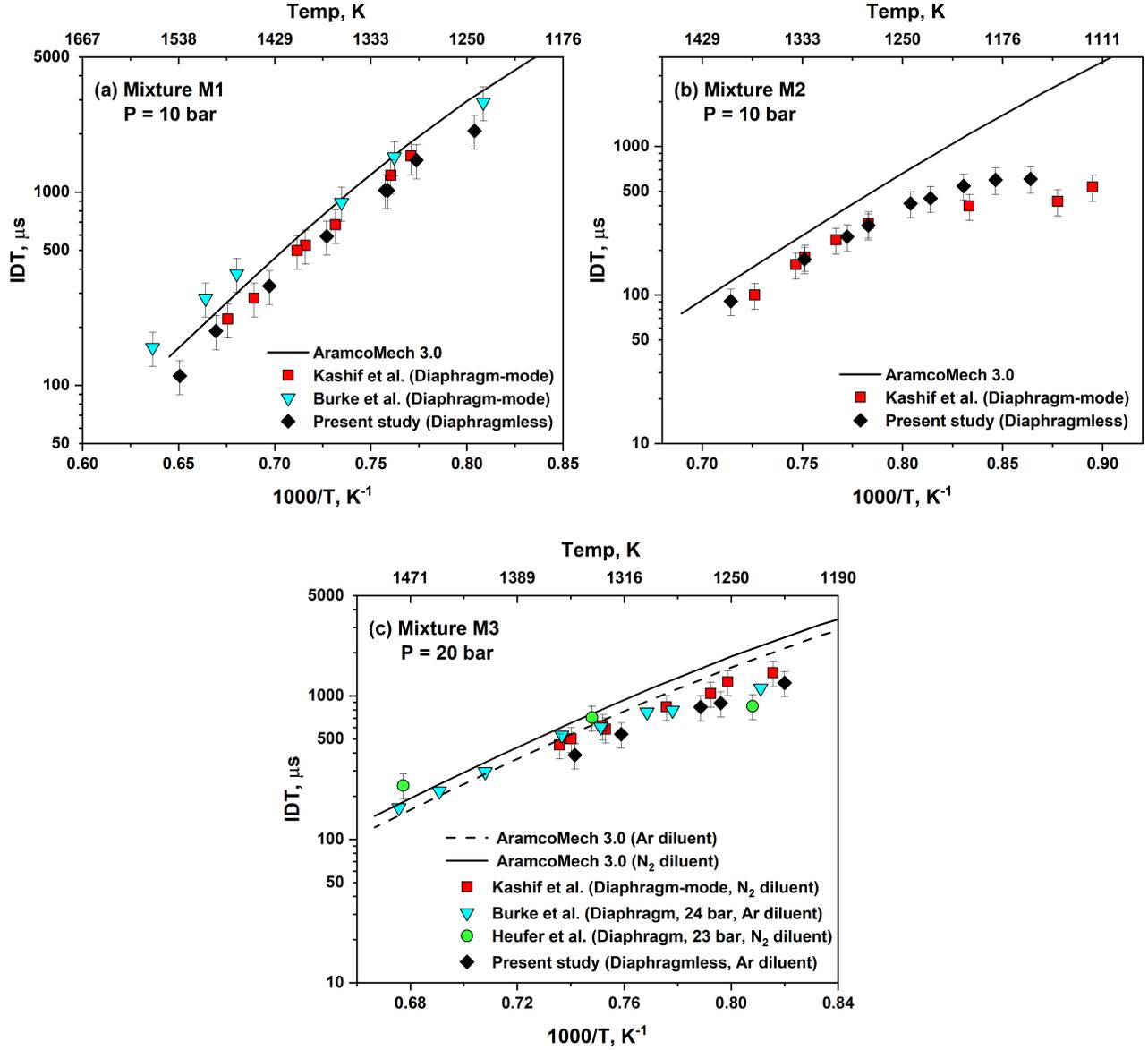

Figure 8. Comparison of IDT measurements in the diaphragmless- and diaphragm-mode for methane mixtures (a) M1, (b) M2, and (c) M3.

The high fraction (90.5%) of diatomic oxygen in mixture M2 makes it a rather non-ideal driven gas, and such a mixture can be prone to premature ignition [37]. The process of shock-ignition is dependent on both space and time in the absence of any dissipative effects [38, 39]. For undiluted mixtures like M2, these effects may play an important role and therefore may be prone to uncertainties at low post-shock temperatures. The shock bifurcation effect is also a possible reason for the deviation in the experimental data compared to the kinetic model. The influence of the boundary layer can be quantified in terms of the bifurcation Damhkohler number ($Da_{BIF}$), defined as [40],

$$Da_{BIF} = \frac{\tau_{BIF}}{\tau_{IDT}}$$

where $\tau_{BIF}$ is the bifurcation time scale and $\tau_{IDT}$ is the ignition time scale. If $Da_{BIF} > 1$, the kinetics of the mixture is unaffected by the physical processes of shock bifurcation. Conversely, when $Da_{BIF} < 1$, the bifurcation growth and the associated local hot spots can ignite the mixture earlier than anticipated. The bifurcation time scale is given by,

$$\tau_{BIF} = \frac{D}{4U_{S2}h'}$$

where $h'$ is the bifurcation height, $D$ is the diameter of the shock tube and $U_{S2}$ is the reflected shock velocity. The bifurcation height is given by the empirical relation,

$$h' = 0.104 M_{S1}^{1.07} \gamma^{-2.66} \left(\frac{W}{W_{O_2}}\right)^{-0.37}$$

where $M_{S1}$ is the incident shock Mach number, $\gamma$ is the specific heat ratio, $W$ is the molecular weight of the mixture and $W_{O_2}$ is the molecular weight of oxygen. Table 2 shows the estimated $\tau_{BIF}$ and $Da_{BIF}$ for the corresponding experimental points obtained using the diaphragmless shock tube. As $Da_{BIF}$ decreases below 1, IDTs are shortened in the experiments due to the chemical kinetics being affected by the shock bifurcation.

Table 2. Estimated bifurcation timescales and Damkohler number for the experimental data points obtained using mixture M2 in the diaphragmless shock tube.

| $T_5$, K | IDT, μs | $\tau_{BIF}$, μs | $Da_{BIF}$ |
|---|---|---|---|
| 1400 | 91 | 360 | 3.95 |
| 1331 | 174 | 382 | 2.20 |
| 1295 | 247 | 395 | 1.60 |
| 1277 | 294 | 402 | 1.37 |
| 1244 | 413 | 414 | 1.00 |
| 1228 | 448 | 420 | 0.94 |
| 1204 | 543 | 431 | 0.79 |
| 1181 | 596 | 440 | 0.74 |
| 1157 | 605 | 451 | 0.75 |

### 3.3. IDTs of n-hexane mixtures

A major distinguishing factor between methane and n-hexane is that n-hexane mixtures (H1 and H2) are expected to exhibit NTC characteristics at intermediate temperatures. Our objective here is to assess the performance of diaphragmless shock tube for studying NTC-type fuels. Figure 9 compares the performance of the diaphragm-type and diaphragmless shock tubes for n-hexane mixtures. Simulations were performed using the Zhang et al. [2] model assuming a *dP/dt* = 2.5 %/ms. For mixture H1, the driver gas was tailored using nitrogen gas (18% by volume) to obtain longer test times. Two separate inlet gas ports were used to fill the gases in the driver section. Nitrogen was filled to the required pressure through an inlet port located at the mid-length of the driver section and, subsequently, helium was filled through a port near the fast-acting valve. Tailoring of the driver gas aided in obtaining IDTs up to 3655 μs. This demonstration shows that the driver gas tailoring method developed for the diaphragm-mode of operation can be applied in the diaphragm-less mode of operation as well. Experimental results reported by Zhang et al. [2] are also included in Figure 9 for comparison. The results obtained using the diaphragmless shock tubes for mixture H1 are in good agreement with those reported by Figueroa-Labastida et al. [31] and Zhang et al. (see Figure 9a). The experiments in the diaphragmless shock tube with fuel-rich mixture H2 also show a good match with the data reported by Figueroa-Labastida et al. [31], as seen in Figure 9b. Predictions using Zhang et al. mechanism shows minor deviations from the experimental data in the temperature range of 714 – 1000 K for mixture H1 (except around 833 K) and in the range of 742 – 833 K for mixture H2. These deviations cannot be solely explained in terms of the bifurcation Damkohler number as done previously for the methane mixture M2. Ignition properties of fuels exhibiting NTC behavior are sensitive to the cold near-wall test gas due to the boundary layer[40]. Accounting for these effects in the simulations may help in capturing the trends obtained in the experiments. Alternatively, the differences between the experimental data and modeling results could be due to the uncertainties in the reaction rate coefficients employed in such large chemical kinetic mechanisms. There is a significant difference in IDTs for mixtures H1 and H2 at similar temperatures due to the different mixture compositions. Mixture H2 (Fuel: 5.00%, $O_2$: 23.75%, Ar: 71.25%) contains more than twice the amount of fuel as compared to mixture H1 (Fuel: 2.20%, $O_2$: 20.50%, Ar: 77.30%). Additionally, mixture H1 (ratio of $N_2$ and $O_2$ is 3.76) is being ignited at air conditions while mixture H2 is oxygen-rich (ratio of $N_2$ and $O_2$ is 3). Therefore, mixture H2 is expected to be more reactive than mixture H1.

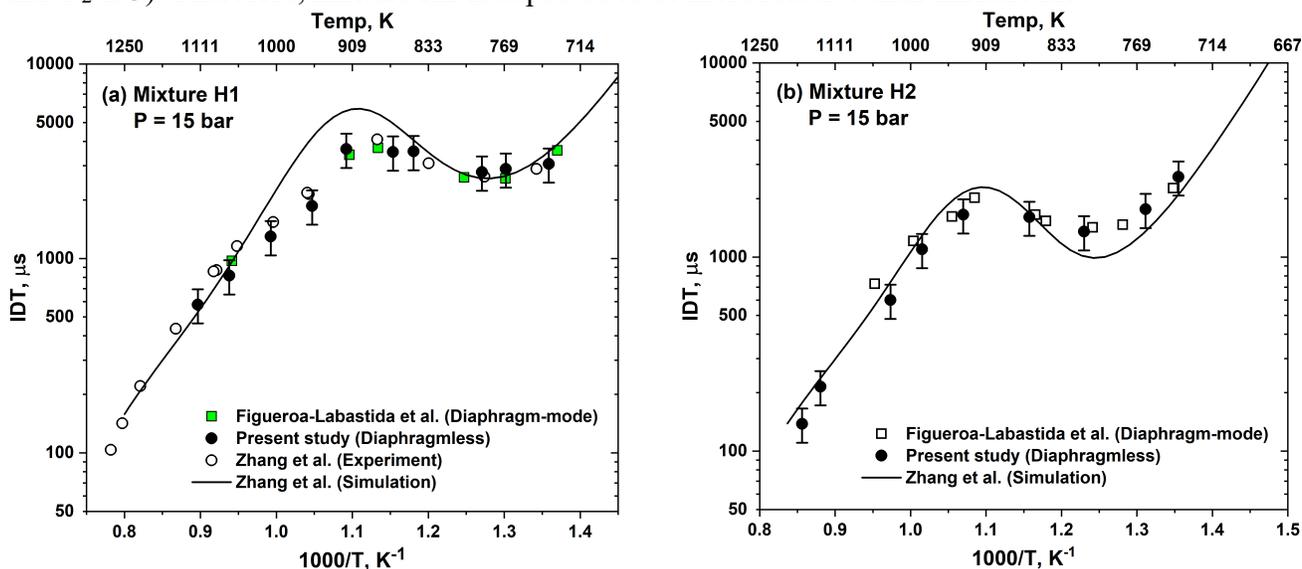

Figure 9. Comparison of IDT measurements in the diaphragmless- and diaphragm-mode for n-hexane mixtures (a) H1, and (b) H2.

## 4. Conclusions

This work describes the development of a diaphragmless shock tube that incorporates a fast-acting valve to replace diaphragms in a conventional shock tube. The diaphragmless shock tube was operated up to driver pressures of about 50 bar and demonstrated good reproducibility of shock conditions. Driver gas tailoring was also implemented to obtain longer test times. IDT measurements in methane and n-hexane mixtures were performed using the diaphragmless shock tube at $P_5$ = 10 - 20 bar and $T_5$ = 738 - 1537 K. The results obtained for fuel-rich, fuel-lean, and oxygen-rich (undiluted) mixtures showed very good agreement with experimental data obtained in the diaphragm-mode and well-validated kinetic models. The proposed diaphragmless shock tube presents many advantages over the conventional shock tubes, such as eliminating the use of diaphragms, avoiding substantial manual effort during experiments, automating the shock tube facility, having good control over driver conditions, and obtaining good repeatability for reliable gas-phase chemical kinetic studies. Future experiments will target driver pressures of greater than 50 bar to obtain higher $P_5$ and $T_5$ conditions. A model to predict the required initial conditions to obtain the desired shock conditions in the diaphragmless shock tube is also planned as a part of the future work.

## Acknowledgements
The work of authors was funded by the baseline research funds at King Abdullah University of Science and Technology.